\documentclass[12pt,article]{aastex}

\usepackage{graphicx,natbib,textcomp,caption,subcaption}         
\usepackage[utf8]{inputenc}
\usepackage{xcolor,amsmath}
\usepackage{multirow}
\usepackage{float}

\slugcomment{ApJ manuscript, \today}

\shorttitle{Dynamic Coronal Seismology}
\shortauthors{Magyar, \& Van Doorsselaere}

\begin{document}

\title{Assessing the capabilities of Dynamic Coronal Seismology of Alfv\'enic waves, through forward-modeling }

\author{N. Magyar\thanks{FWO (Fonds Wetenschappelijk Onderzoek) PhD fellow}, T. 
Van Doorsselaere}
\affil{Centre for mathematical Plasma Astrophysics (CmPA), KU 
Leuven, Celestijnenlaan 
200B bus 2400, 3001 Leuven, Belgium; norbert.magyar@wis.kuleuven.be}

\begin{abstract}

Coronal seismology is a diagnostic tool used in solar physics, for measuring parameters which otherwise are hard to measure, arguably most importantly magnetic field values. The parameters are inferred by combining observations of waves and magnetohydrodynamic (MHD) wave theory. As of now, coronal seismology was successfully applied to various single oscillation events. Such events are relatively rare, resulting in rare occasions to apply diagnostics. Ubiquitous waves in the solar atmosphere might however allow for the possibility of dynamic coronal seismology, continuous inversions for coronal parameters, which would constitute a huge leap forward in many areas of solar physics. In this paper, we investigate the robustness and accuracy of magnetic field diagnostics applied to forward-modeled three-dimensional (3D) MHD simulations of propagating Alfv\'enic waves. We find that the seismologically measured magnetic field values are reassuringly close to the input value (within $\approx 20\%$) for a range of setups studied, providing encouragement and confidence for the further development of dynamic coronal seismology.
\end{abstract}


\section{Introduction}

The solar corona is still enigmatic from a scientific point of view, with important unsolved questions about its nature pending, such as solar wind acceleration and coronal heating
\citep{ROG:ROG1641,2012RSPTA.370.3217P}. While finding solutions to these questions is important on its own, it is expected to have major implications in related fields, such as space weather \citep{2010SGeo...31..581S}, which is set to grow in importance as we make advances in technology and space exploration. One impediment towards solving coronal mysteries is the notorious difficulty in determining its key physical parameters, such as the magnetic field, by direct spectroscopic or polarimetric measurements. Early observational evidences of waves in coronal structures using SOHO/TRACE \citep{1999SoPh..186..207B,1999ApJ...520..880A} opened the way for the previously theorized coronal seismology \citep{1970PASJ...22..341U,1984ApJ...279..857R}, as a tool for coronal plasma diagnostics. The first attempt to sesimologically determine the magnetic field of transversely oscillating coronal loops was applied to the first such event observed by \citet{2001A&A...372L..53N}. Since then, coronal seismology was successfully applied to numerous oscillation events \citep[for reviews see, e.g.][]{2005RSPTA.363.2743D,2012scsd.book.....S,2012RSPTA.370.3193D}.
A common feature of all previously diagnosed events is their localization in time, i.e. single events. This obviously limits the applicability of seismology to the brief duration of the oscillation event. Moreover, these single oscillation events tend to be rare, as they are mostly related to flaring events or eruptions, meaning that seismology is restricted to brief diagnostics highly localized in time, or in time and space for non-global oscillations. However, the discovery of ubiquitous propagating transverse oscillations with CoMP \citep{2007Sci...317.1192T,2009ApJ...697.1384T} and SDO \citep{2011Natur.475..477M} or the more recently identified nearly ubiquitous, decay-less low-amplitude kink coronal loop oscillations \citep{2013A&A...552A..57N,2013A&A...560A.107A,2015A&A...583A.136A}, as well as oscillations in plumes \citep{2014ApJ...790L...2T} lead to the possibility of continuous diagnostics in time, i.e. dynamic coronal seismology. Consequently, very recently the first seismologic `magnetic field image' was obtained, based on the ubiquitous transverse waves \citep{2017A&A...603A.101L}, using a methodology put forth by \citet{2015NatCo...6E7813M}. In this study, the authors use the MHD kink phase speed of a flux tube \citep{1983SoPh...88..179E} as the inversion tool for the magnetic field. The ubiquitous propagating transverse waves are now widely regarded as Alfv\'enic waves \citep{2012ApJ...753..111G}, although this reinterpretation of the nature of the waves does not modify the phase speed formula used in the inversion. However, it is still assumed that the observed ubiquitous waves are transverse oscillations of flux tubes, and while the fine structure in the corona is still unknown \citep{2013A&A...556A.104P,2014LRSP...11....4R,2017ApJ...840....4A}, it is unlikely that the magnetic cylinder model is satisfactory. 
Structuring across the magnetic field, among other factors \citep{2009ApJ...699L..72D,2014ApJ...784..101P}, is an important detail for seismology, as it can greatly influence the nature and propagation of MHD waves \citep{2010ApJ...714.1637V,2010A&A...524A..23T,2010ApJ...716.1371L}, that is, altering the dependence of the observed phase speed on physical properties such as magnetic field or mass density, \citep[e.g.][]{2007A&A...475..341V,2013ApJ...769L..34A}, which might lead to erroneous inversions if not considered. Some light was shed recently on the weaknesses of modeling the corona as a bundle of independent thin magnetic strands by \cite{2016ApJ...823...82M}. In their simulations, a loop consisting of packed strands is quickly deformed and mixed when disturbed by propagating transverse waves, leading to a turbulent cross-section \citep{2017NatSR...14820M}. This result reiterates the above mentioned need to move away from rigid cylindrical models, in favour of more realistic descriptions, accounting for nonlinearities. \par
In this paper, we attempt to determine the accuracy of inferring the magnetic field of a simulated column of coronal plasma permeated by ubiquitous transverse waves, by means of coronal seismology. We explore the effect of several different parameters, such as density distribution in the cross-section and density ratio, isothermal vs. non-isothermal models, resolution and cadence, and the influence of the chosen spectroscopic line on the results of the inversion. \par
This study appears to be very timely, as in the near future, new powerful ground-based observatories, namely the 4 m Daniel K. Inouye Solar Telescope (DKIST), 2 m Indian National Large Solar Telescope (NLST), and the proposed 1.5 m Coronal Solar Magnetism Observatory \citep[COSMO,][]{2016JGRA..121.7470T} will provide full-disk polarimetric observations of the corona at unprecedented resolution and cadence, ideal for applying dynamic coronal seismology to.

\section{Numerical model and methods}

\subsection{Method and mesh}

We run three-dimensional (3D), ideal MHD simulations using the finite-volume \texttt{MPI-AMRVAC} code \citep{2012JCoPh.231..718K,2014ApJS..214....4P}, opting for the implemented one-step \texttt{TVD} method with Roe solver and `Woodward' slope limiter. The ideal MHD equations are supplemented with the ideal gas law for a plasma with mean mass per particle $\approx 0.83$ of the proton mass, for coronal abundances. The zero-divergence condition of the magnetic field is enforced via Powell's scheme. Our numerical domain is an elongated box, aiming to represent a thin section of the corona, measuring $5 \times 5 \times 50\ \mathrm{Mm}$, covered by $256^2 \times 128$ numerical cells. Thus the resolution is much higher in the $x-y$ plane perpendicular to the direction of the mean magnetic field, than in the direction parallel to it ($z$-axis), along which we expect the solution to be smooth. The total runtime of the simulations is $t_f = 1000\ \mathrm{s}$.

\subsection{Initial conditions}
\label{initcond}
The plasma is in equilibrium initially. As we explore the parameter space, we employ various initial conditions for different runs. The plasma density inside the domain is described by:
 \begin{equation}
  \begin{array}{lr}
   \rho(x,y,z) = \rho_0 + \sum\limits_{i=1}^{N} A_{i} \exp^{-[(x-x_{i})^2 + (y-y_{i})^2]/ \sigma_{i}^2},
  \end{array}
 \label{Gaussiandens}
\end{equation}
where $\rho_0 = 2 \cdot 10^{-13}\ \mathrm{kg\ m^{-3}}$ is the background density, N is the number of Gaussian density enhancements added. Varying N we study the effect of different filling factors. Furthermore, $x_i, y_i, A_i, \sigma_i$ is the random (from uniform distributions) $x,y$ position, amplitude, and width, respectively, of the $i$\textsuperscript{th} enhancement. The limits of the uniform distributions are the box size $[-2.5\ \mathrm{Mm},2.5\ \mathrm{Mm}]$ for $x_i$ and $y_i$, $[0,A_{max}]$ for $A_i$, and $[0,3.5 \cdot 10^{-1}\ \mathrm{Mm}]$ for $\sigma_i$. Varying the $A_{max}$ parameter, we explore the effects of the density ratio (see first row of Fig.~\ref{dynamics}). For the other runs in this paper, $A_{max} = 5 \rho_0$. As there is no consensus on the average width of strands which compose coronal structures \citep[see, e.g.][]{2013A&A...556A.104P,2013ApJ...772L..19B}, the choice for the range limit was somewhat arbitrary. Nevertheless, in the light of the recent findings in \citet{2016ApJ...823...82M}, we expect structures on a large range of scales, continuously deforming when exposed to propagating Alfv\'enic waves, as shown in Fig.~\ref{dynamics} \par

For most of the runs, we employ a uniform and straight magnetic field of $5\ \mathrm{G}$, directed along the $z$-axis. In this case, plasma $\beta = 0.1$, plasma (thermal) pressure is constant throughout the domain, and the temperature varies according to the ideal gas law. For example, in the multithermal $N=50$ setup, the plasma density varies in the range $\approx [\rho_0,10\rho_0]$, and the temperature varies in the range $\approx [0.45\ \mathrm{MK},4.3\ \mathrm{MK}]$. \par To study the influence of multithermal versus isothermal plasmas, we also employ models with a constant temperature of $1.03\ \mathrm{MK}$, while keeping the density structure the same as in the corresponding multithermal setup. In this case, the plasma pressure is calculated from the ideal gas law, and the magnetic field from the constant total pressure ($p_T = p + \frac{B_0^2}{2\mu_0} \approx 0.1\ \mathrm{Pa}$) equilibrium condition, leading to an inhomogeneous magnetic field with the same average value as the uniform field. \par
Note that the equilibrium is independent of the $z$-axis: there is no stratification, implying that we neglect gravity. However, stratification has important effects on wave  propagation: it leads to a height-dependent phase speed and oscillation amplitude \citep[e.g.][]{2011ApJ...736...10S}. Including stratification would allow for height-dependent inversions, but we do not expect it to have an effect on the accuracy of the inversions, as the phase speed determination can be `localized' along a short segment (see subsection~\ref{phasespeed}) when compared to the typical density scale heights in the solar corona ($\approx 50\ \mathrm{Mm}$).  We also neglect sources or sinks in the energy equation: thermal conduction, radiative losses, and heating terms. In this sense, we are limited to the study of the effect of transverse structuring.

\subsection{Wave driver}

Alfv\'enic waves are driven at the $z = 0$ (bottom) boundary. We intend to mimic the properties of the observed propagating transverse waves in coronal structures \citep{2015NatCo...6E7813M}. In this sense, we employ a driver of the following form:
 \begin{equation}
  \begin{array}{lr}
   \mathbf{v}_x(x,y,0,t) = \sum\limits_{i=1}^{10} U_{i} \sin(\omega_i t), \\
   \mathbf{v}_y(x,y,0,t) = \sum\limits_{i=1}^{10} V_{i} \sin(\omega_i t), 
  \end{array}
 \label{driver}
\end{equation}
where $U_i$ and $V_i$ are random velocity amplitudes from a uniform distribution with limits $[-10\frac{1}{\sqrt{2}}\ \mathrm{km\ s^{-1}},10\frac{1}{\sqrt{2}}\ \mathrm{km\ s^{-1}}]$, and $\omega_i = \frac{2 \pi}{T_i}$ are random angular frequencies from the observed log-normal distributions of wave periods \citep{2015NatCo...6E7813M}. Therefore, we launch randomly polarized transverse waves. The resulting root mean-square velocity amplitude is $\approx 12\ \mathrm{km\ s^{-1}}$, close to the observed value. Note that the driver is time but not space-dependent: the whole bottom boundary is driven with the same velocity. 

\subsection{Boundary conditions}

At the bottom boundary, in addition to the previously described conditions for the velocity, we use Neumann-type zero-gradient `open' conditions for all the other variables. Open boundary conditions are used at the $z = 50\ \mathrm{Mm}$ (top) boundary as well, allowing waves to leave the domain with minimal reflections. Laterally, we use periodic boundary conditions. In this sense, to ensure that the variables are continuous at the side boundaries initially, we continue periodically the density enhancements with respect to the closest $x$ and $y$-axis boundaries. This is done by setting $x_i$ and $y_i$ of Eq.~\ref{Gaussiandens} as multi-valued, conditional parameters:
 \begin{equation}
    x_i = \begin{cases}
    \lbrace x_i,x_i-L \rbrace,&if\ x_i \geq 0\\
    \lbrace x_i,x_i+L \rbrace,&if\ x_i < 0
\end{cases} \\
 \end{equation}
and similarly for $y_i$. Here $L = 5\ \rm{Mm}$ is the width of the square cross-sectional numerical domain. Continuing periodically on the closest boundaries is sufficient as only enhancements close to the boundaries are significant in this sense.

\begin{figure*}[h!]
    \centering
       \begin{tabular}{@{}cc@{}}
        \includegraphics[width=0.5\textwidth]{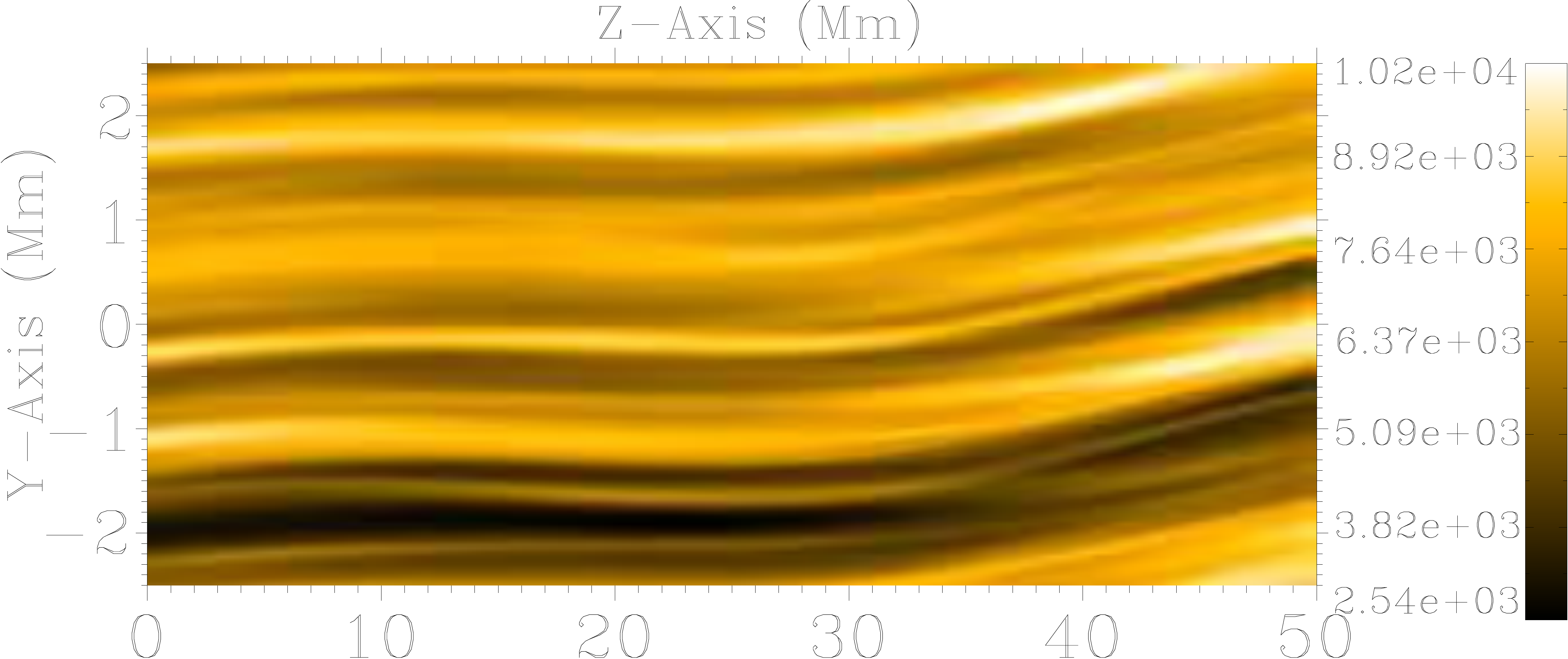}  
        \includegraphics[width=0.5\textwidth]{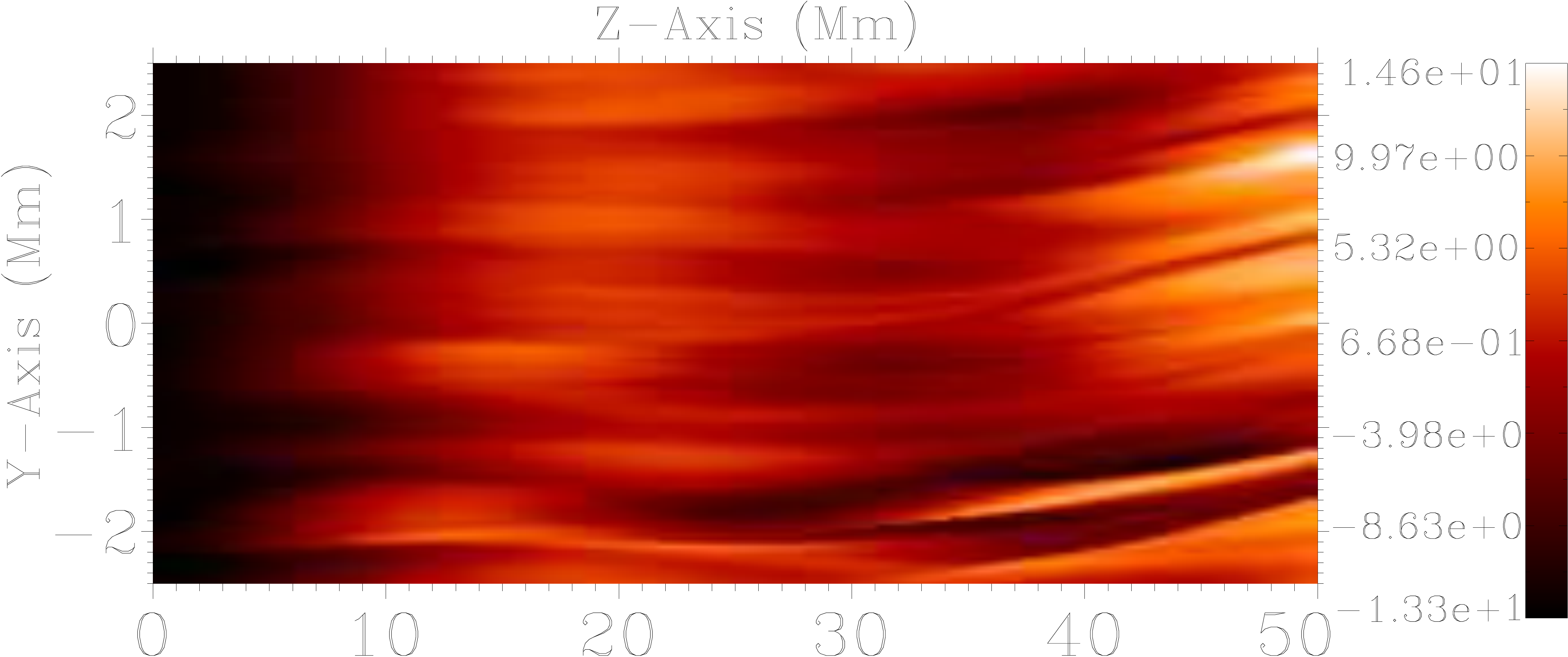} \\
       \end{tabular}  
        \caption{Example of forward-modeled synthetic images, for the setup with $N=250$, at $t=500\ \mathrm{s}$, using the Fe IX 171{\AA} line with LOS perpendicular to the $y-z$ plane. \textit{Left:} Intensity image with arbitrary units. \textit{Right:} Doppler shifts in units of $\mathrm{km\ s^{-1}}$. (In the online version of the paper, the two images are animated, representing their evolution from $t_0 = 0$ to $t_f = 1000\ \mathrm{s}$).}
        \label{fomo}
 \end{figure*}
 
\subsection{Forward-modeling and phase speed determination}
\label{phasespeed}

For a more realistic study, it is essential that we convert the simulation results to synthetic (spectroscopic) images. This is done by using \texttt{FoMo} \citep{10.3389/fspas.2016.00004}. By using forward-modeling, we can apply the same techniques for the analysis of simulation results as used for real observational data. We convert our datacubes to spectroscopic (i.e. calculating intensity, Doppler shifts and spectral widths) images (see Fig.~\ref{fomo}) using several spectral lines: Fe XII 186.887{\AA}, 193.509{\AA}, 195.119{\AA}, and Fe IX 171.073{\AA}. These specific lines were chosen for their frequent use in observational studies. \par

\begin{figure*}[h!]
    \centering
       \begin{tabular}{@{}cc@{}}
        \includegraphics[width=0.4\textwidth]{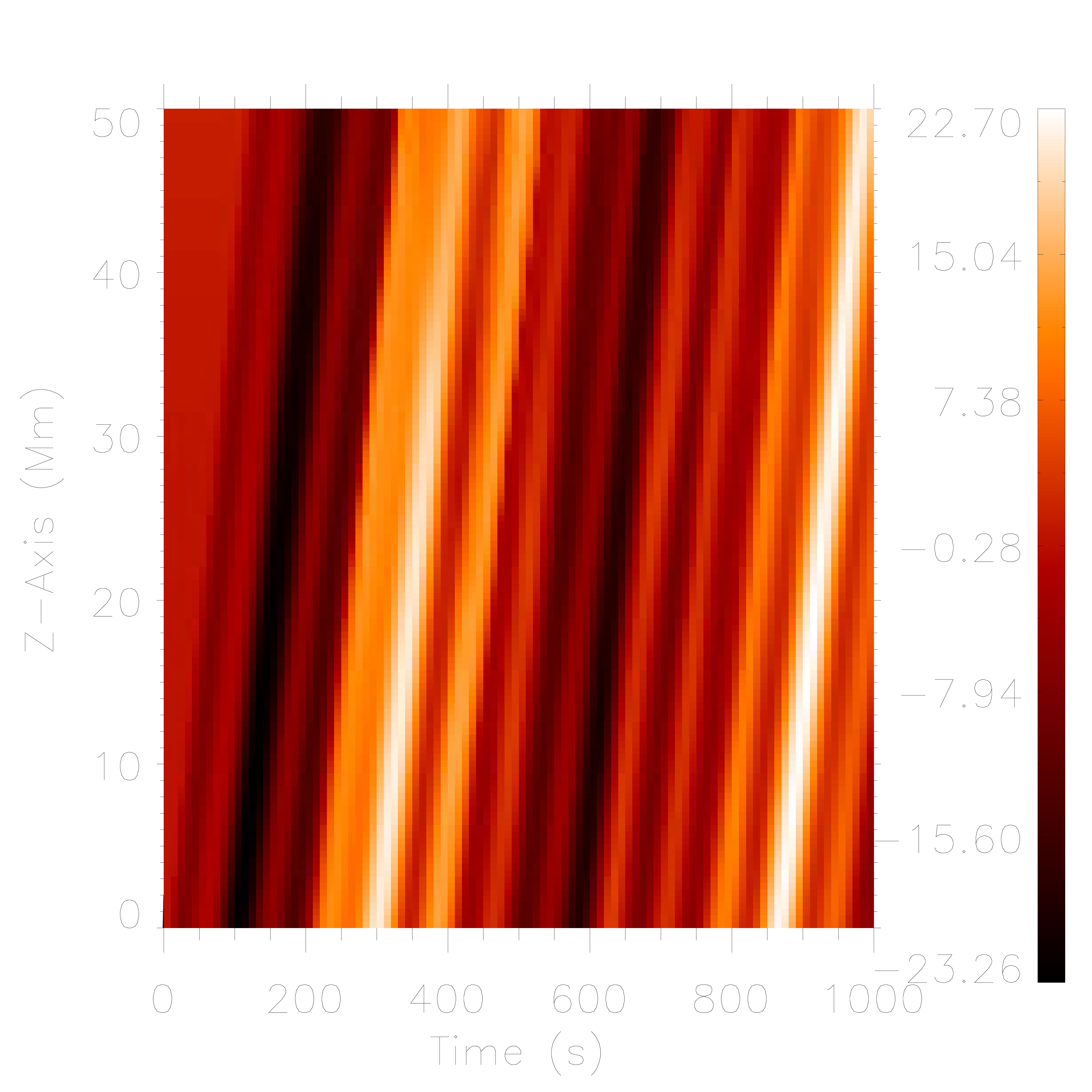}  
        \includegraphics[width=0.37\textwidth]{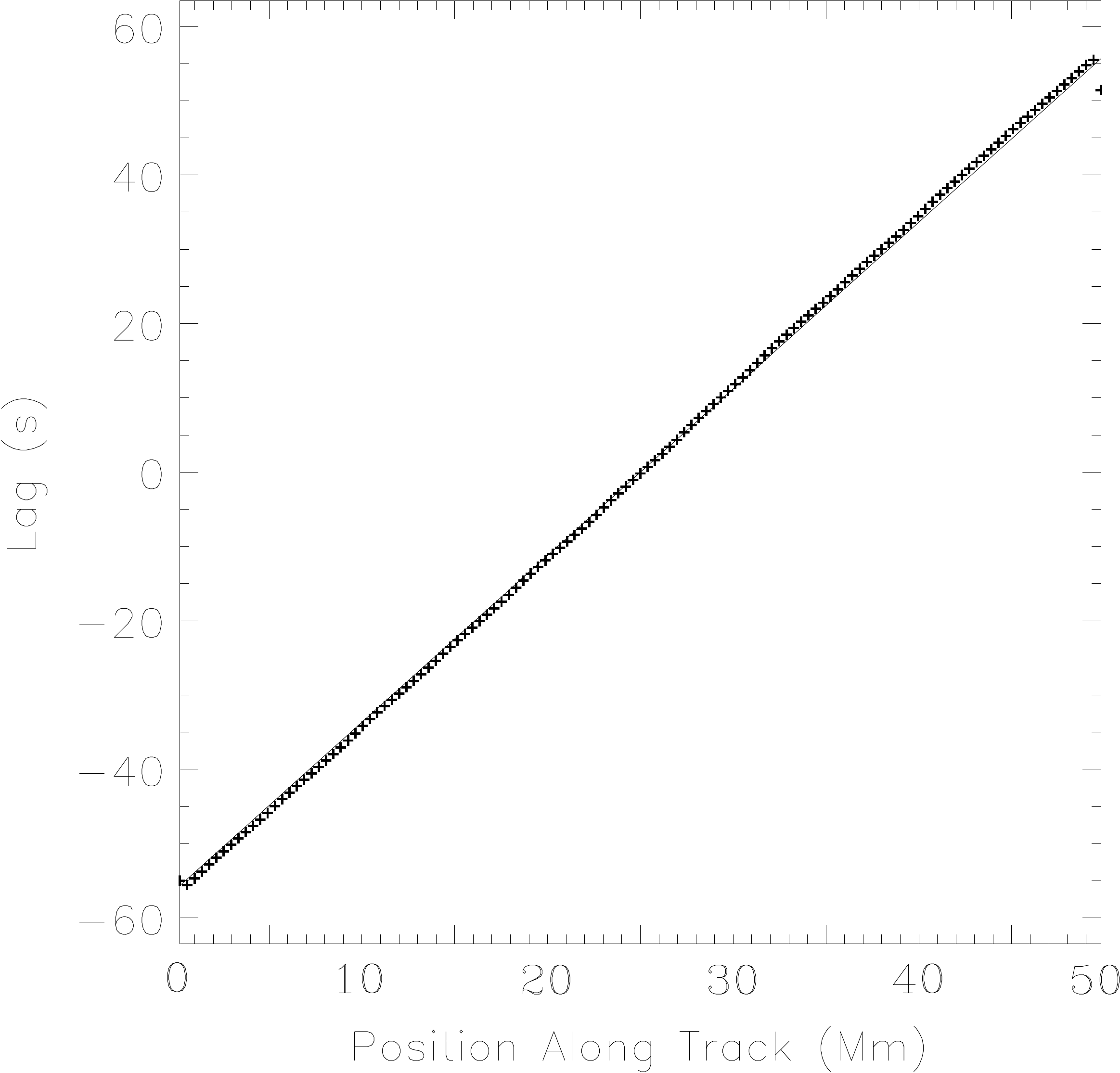} \\
       \end{tabular}  
        \caption{\textit{Left:} Time-distance diagram of the Doppler velocity, along the $x=0,y=0$ line, for the $N=250$ run. Velocity is shown in units of $\mathrm{km\ s^{-1}}$. \textit{Right:} Result of cross-correlating the time-series at the centre of the $z$-axis: The slope of the observed lag along the track is the phase speed of the waves.}
        \label{corr}
 \end{figure*}
 
The phase speed of the propagating waves is used for the seismologic determination of the magnetic field. For measuring the phase speed we employ a technique used previously for observations, as described in \cite{2009ApJ...697.1384T}. It is based on cross-correlating Doppler time series, which result from the forward-modeling of the simulations. More specifically, the phase speed is determined along so-called `wave-paths' by cross-correlating the time series at the centre of the path with those across the entire
length of the path (see Fig.~\ref{corr}). The wave-paths are line segments in the direction of wave propagation, in our case the $z$-axis. We choose the line segment to span the whole length of the $z$-axis ($50$ Mm), however tests show that accurate phase speed measurements are possible for shorter segments as well, down to 9 numerical cells in length ($\approx 3.5\ \mathrm{Mm}$). For obtaining the phase speeds, we use the forward-modeled Fe XII 195.119{\AA} Doppler shift time series. A study on the influence of the chosen line is made in subsection~\ref{spectroline}. 
 
\subsection{Density diagnostics}

\begin{figure}[h!]
    \centering
        \includegraphics[width=0.4\textwidth]{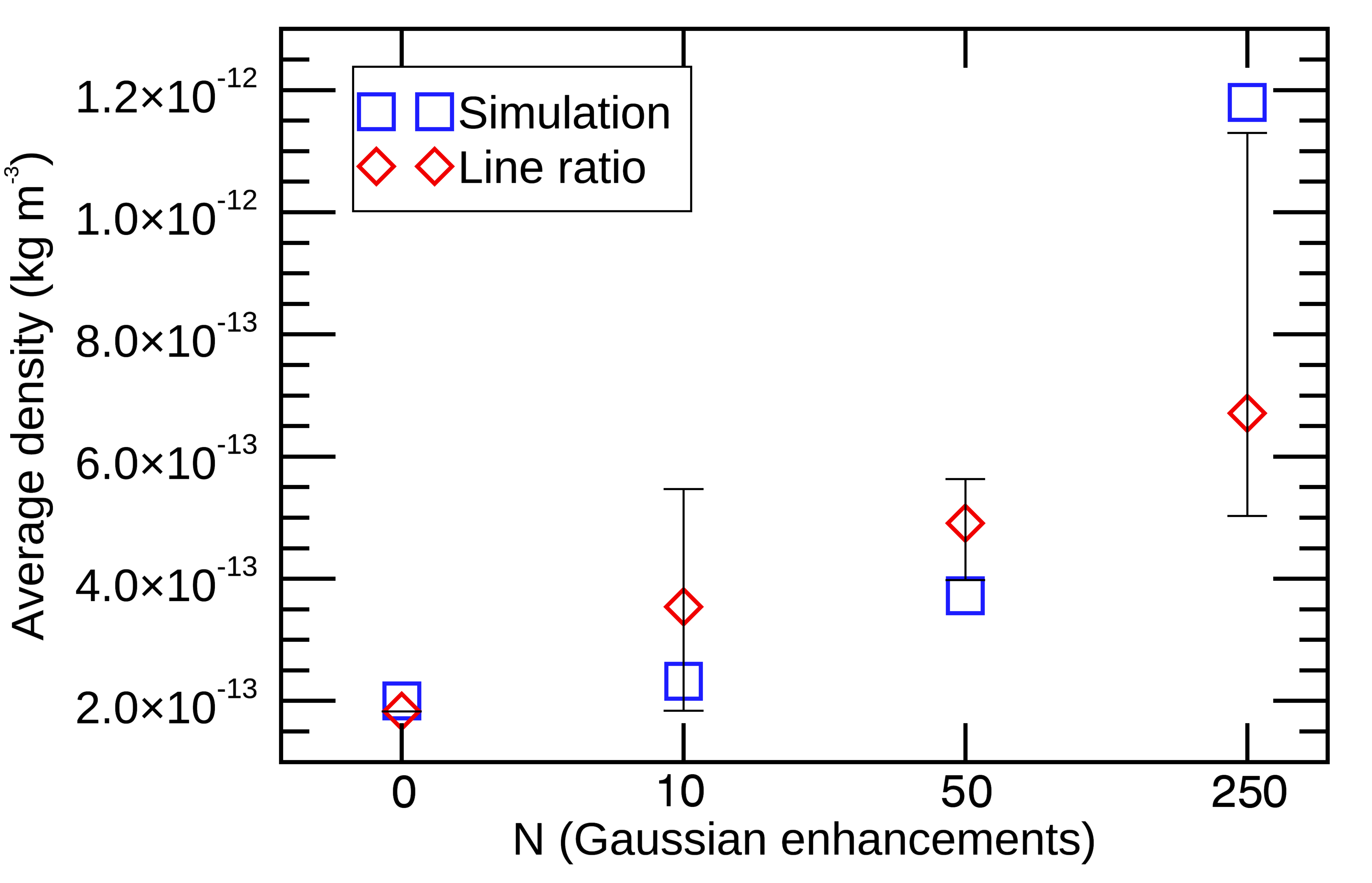}  
        \caption{Comparison between the (volume) average mass density used as an initial condition in the simulation, and the average density determined by the line ratio method from the forward-modeled images, using the $y-z$ plane as the plane-of-sky (POS). The error bars extend from the minimum to the maximum density value found in this plane.}
        \label{densdiag}
\end{figure}

A pair of emission lines emitted from the same ion for which the ratio of the emissivities is dependent on the electron density of the plasma \citep{1997A&A...327.1230L} can be used as a density diagnostic. We use the lines Si X  261.056{\AA} and 258.374{\AA} to determine the density in the forward-modeled simulations at $t=0\ \mathrm{s}$, using \texttt{eis\char`_density} from the \texttt{SSW} package. Opting for this particular line ratio was motivated by its good sensitivity in the range of average densities present in our simulations ($\approx 10^8 - 10^9\  \mathrm{cm^{-3}}$ ), and its relative insensitivity to plasma temperature, which is beneficial for our multithermal setups. By doing this, we obtain a density value for each pixel of the forward-modeled images. For inhomogeneous setups $(N \neq 0)$ we average over these pixels to obtain an average density. The results of this process, in comparison with the simulation input, can be seen in Fig.~\ref{densdiag}. For the homogeneous case ($N=0$) the measured density and the density in the simulation are close, and deviations can be explained based on the aforementioned temperature dependence of the line ratio. However, for the inhomogeneous cases, deviations from the input value appear. This is the result of an interplay between the density and temperature inhomogeneities and the density and temperature-dependent contribution function ($G_\lambda(n,T)$) of a specific line in an optically thin plasma. The intensity of a pixel is proportional to the integral along the LOS \citep[e.g.][]{10.3389/fspas.2016.00004}:
\begin{equation}
  I \propto \int n^2 G_\lambda(n,T) dh,
 \label{goft}
\end{equation}
so that most of the contribution will come from structures for which the emissivity is the highest. In this sense, it is these structures that define the measured line ratio, and ultimately the density in a specific pixel.
This leads to an overestimation of the average density in the ($N =10,50$) setups. Conversely, for the highly inhomogeneous case ($N=250$), the average density is underestimated by a factor of two, as the temperature of the densest structures is far from the peak formation temperature of the lines. For the seismological inversions, we consider the maximum and minimum inferred density values in the 2D density maps as the `error' limits. The difference between the measured and the input density is likely to be dependent on the LOS, as the overlapping of structures is dependent on the point of view. Note that, because of the dynamics of the plasma perturbed by the driven Alfv\'enic waves in our simulations (see Fig.~\ref{dynamics}), the structuring changes rather drastically in the cross-section. This implies that the density determined through the line ratio method is time-dependent. However, we found no significant changes ($ < 10\%$) of the determined average density with time.

\section{Results}
\label{results}
\subsection{Uniform density}

As an initial test-run, a simulation with constant density ($N=0$) was set up. In this case, the driven propagating waves are pure Alfv\'en waves: solely driven by magnetic tension, causing no compression, and propagating at the well-known Alfv\'en speed:

\begin{equation}
 V_\mathrm{A} = \frac{|\mathbf{B_0}|}{\sqrt{\mu_0 \rho_0}}
 \label{Alfven}
\end{equation}

As a side note, although the Alfv\'en waves are not compressional, nonlinearity still causes minute perturbations in pressure and velocity along the magnetic field, unrelated to the slow magnetosonic waves, by the so-called ponderomotive effects \citep[e.g.][]{2013SoPh..288..205T}. \par
The measured phase speed (by the methods listed in the previous section) is $0.996\ \mathrm{Mm\ s^{-1}}$, while the real Alfv\'en speed in the simulation is $0.997\ \mathrm{Mm\ s^{-1}}$, an $\approx 0.14 \%$ error, demonstrating the accuracy of the method in homogeneous conditions. To determine the magnetic field strength, we need to use the density determined by the line ratio method: the seismologically measured value for the magnetic field, by using Eq.~\ref{Alfven}, is then $4.79\ \mathrm{G}$, while the input value is $5\ \mathrm{G}$. Note that this deviation ($\approx 4\%$) is attributable to the slight temperature dependence of the line ratio. This method assumes that the plasma temperature is near the peak formation temperature of the lines (in the case of Si X, $\approx 1.4\ \mathrm{MK}$), while the temperature in the homogeneous simulation is $\approx 4.3\ \mathrm{MK}$.

\subsection{Effect of density structuring}

It is well-known that structuring across the magnetic field, in the general case, leads to the linear coupling of Alfv\'en and magnetosonic waves, resulting in MHD waves with mixed properties, i.e. waves presenting both Alfv\'en and magnetosonic properties, propagating both vorticity and compression \citep{2011SSRv..158..289G}. In this sense, it is generally not possible to classify waves in inhomogeneous plasmas as strictly belonging to one of the three MHD wave types found in an infinite and homogeneous plasma. However, in some situations, as in the case of the observed ubiquitous transverse waves and the waves in the present study, the predominantly Alfv\'en characteristics let us refer to these waves as Alfv\'enic \citep{2009A&A...503..213G,2012ApJ...753..111G,2015NatCo...6E7813M}. In a thin, cylindrical flux tube, Alfv\'enic waves propagate with the kink speed:
\begin{equation}
 C_\mathrm{k} =  V_\mathrm{Ai}\sqrt{\frac{2}{1+\rho_0/\rho_i}}
\end{equation}
where $V_\mathrm{Ai}$ is the Alfv\'en speed inside the tube, and $\rho_i, \rho_0$ are the internal and external densities, respectively.

\begin{table}[h!]
  \caption{\label{avgdens}Results of the seismological determination of the average magnetic field, compared to the input values, as a function of $N$. The measured phase speed is also compared to the input average Alfv\'en speed. One standard deviation is taken as error margin on each side of the measured average Alfv\'en speeds. For the magnetic field estimate, the highest and lowest values computed are taken as error margins, taking into account the Alfv\'en speed and density error margins. The percent error is relative to the corresponding input values. }
  	\centering
  	\begin{tabular}{|l|c|c|c|c|c|c}
  		\hline\hline
  		\textit{N} & $\langle {V_A}_{seism} \rangle (\mathrm{km\ s^{-1}})$ & $ \langle {V_A}_{input} \rangle (\mathrm{km\ s^{-1}}) $ & $\langle |\mathbf{B}_{seism}| \rangle \ (\mathrm{G})$ & $ |\mathbf{B}_{input}|\ (\mathrm{G})$ \\ 
  		\hline
  		0 & $996$ (0.1\%)& 997 & $4.79$ (4.2\%)& \multirow{4}{0em}{5} \\
  		10 & $877 \pm 60$ (5.5\%) & 928 & $5.86\substack{+1.9\\-1.9}$ (17.2\%)&\\
  		50 & $649 \pm 30$ (11.2\%)& 731 & $5.14\substack{+0.6 \\ -0.7}$ (2.8\%)&\\
  		250 & $498 \pm 14$ (21.2\%)& 411 & $4.52\substack{+1.5 \\ -0.7}$ (9.6\%)&\\
  		\hline
  	\end{tabular}
  	\label{table}
\end{table}

However, for a generalized structuring across the magnetic field, as in the present case, and very likely in reality, it is impossible to derive such a simple relation for the phase speed. In studies with multiple interacting thin flux tubes \citep[e.g.][]{2008A&A...485..849V,2008ApJ...676..717L,2011ApJ...731...73P}, also called the multi-stranded model, \cite{2008ApJ...679.1611T} showed that global-mode oscillations are present, and suggested that the internal structuring of flux tubes might not change much the global oscillation properties. This was somewhat contradicted by \cite{2010ApJ...716.1371L} who found no global modes through normal mode analysis, but instead a broad band of collective modes, with complicated individual strand behaviour. Furthermore, \cite{2016ApJ...823...82M} showed that, when perturbed by propagating Alfv\'enic waves, multi-stranded loops tend to mix and deform, therefore continuously changing their transverse structuring. Based on the above arguments, it seems a reasonable choice to use the average Alfv\'en speed as the inversion tool for Alfv\'enic waves, as used previously in \citet{2015NatCo...6E7813M,2017A&A...603A.101L}:
\begin{equation}
 \langle V_\mathrm{A} \rangle = \frac{\langle |\mathbf{B}| \rangle}{\sqrt{\mu_0 \langle \rho \rangle }},
 \label{avgAlfven}
\end{equation}
where by average we mean either in the 3D simulation domain (denoted by the `\textit{input}' subscript in the following), or in the forward-modeled data (denoted by the `\textit{seism}' subscript in the following).  \par
\begin{figure}[h!]
    \centering
        \includegraphics[width=0.7\textwidth]{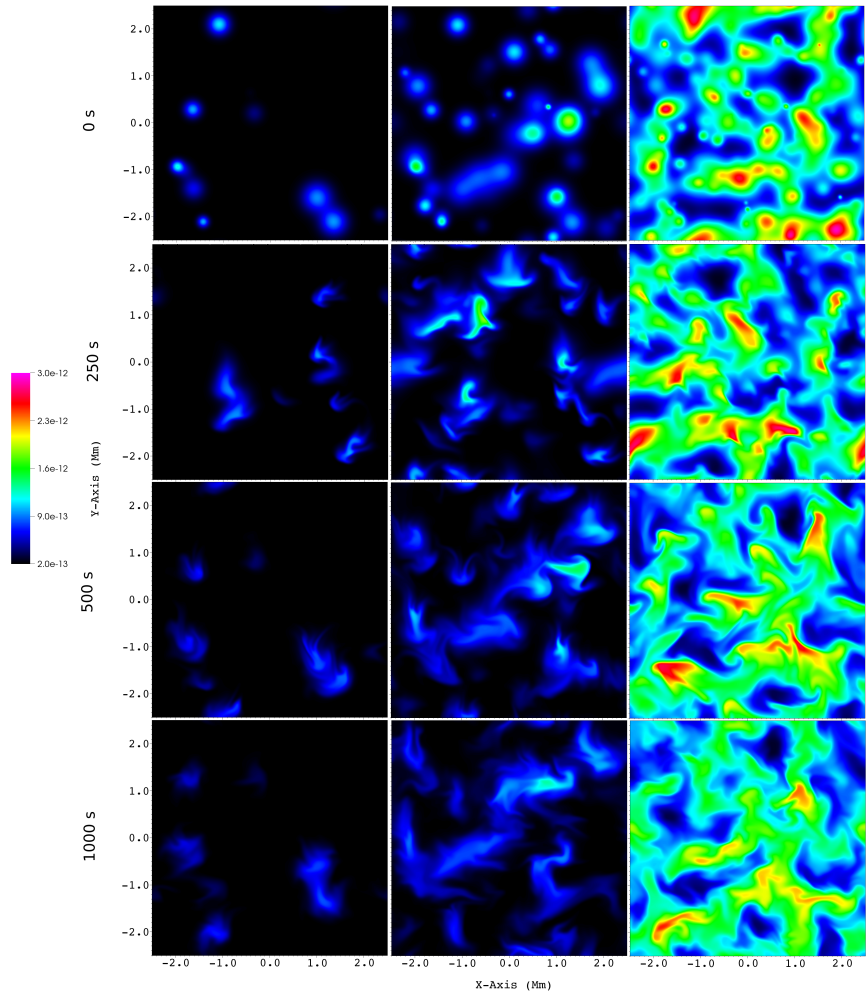}  
        \caption{Evolution of the density structure in the cross-section ($x-y$ plane) at $z = 40\ \mathrm{Mm}$, for $N = 10,50,250$ from left to right for different columns, respectively, at simulation times indicated left of each row.  Density is in units of $\mathrm{kg\ m^{-3}}$.}
        \label{dynamics}
\end{figure}

We determined the average magnetic field value seismologically by using the average Alfv\'en speed as the inversion tool for different values of $N$, the results of which can be found in Table~\ref{avgdens}. Higher values of $N$ reflect a higher degree of inhomogeneity and filling factor (of high-density structures), as shown in Fig.~\ref{dynamics}. Note that the measured value of the phase speed varies slightly as a function of the chosen $y-z$ axis wave-path or line segment on the synthetic images used for the cross-correlation. By averaging we mean the average phase speed over different $y-$axis origins (for the lines, all parallel to the $z-$axis), with one pixel as the increment. The standard deviation of the phase speeds measured is used as error margins.

\subsection{Effect of density ratio}

\begin{table}[h!]
  \caption{Results of the seismological determination of the average magnetic field for different density ratios, for the N=50 case. The meaning of errors and percentages is the same as for Table~\ref{table}.}
  	\centering
  	\begin{tabular}{|c|c|c|c|c|c}
  		\hline\hline
  		$A_{max}$ & $\langle |\mathbf{B}_{seism}| \rangle\ (\mathrm{G})$ & $ |\mathbf{B}_{input}|\ (\mathrm{G})$ \\ 
  		\hline
  		$2\rho_0$ & $5.84\substack{+0.7 \\ -1.4}$ (16.8\%) & \multirow{3}{0em}{5} \\
  		$5\rho_0$ (default)& $5.14\substack{+0.6 \\ -0.7}$ (2.8\%) &\\
  		$15\rho_0$ & $5.14\substack{+0.7 \\ -0.7}$ (2.8\%) &\\
  		\hline
  	\end{tabular}
  	\label{densratio}
\end{table}

Besides the degree of density inhomogeneity, the difference in density between different structures is also an important parameter. We consider the $N = 50$ case, with the same magnetic field strength and density distribution, but we vary $A_{max}$, described in subsection~\ref{initcond}, setting it to a lower ($A_{max} = 2\rho_0$) and a higher ($A_{max} = 15\rho_0$) value for a smaller and higher density ratio, respectively. Interestingly, the simulated line ratio density diagnostic shows an erroneous small variation with the density ratio. We obtain $ \{ 4.28, 4.91, 5.06 \} \cdot 10^{-13}\ \mathrm{kg\ m^{-3}}$ average densities for the three cases (in growing density ratio order), while the real average densities in the simulation are $\{ 2.7, 3.72, 7.17 \} \cdot 10^{-13}\ \mathrm{kg\ m^{-3}}$, respectively. The cause for this failure to differentiate between the different density ratio setups by the line ratio method must lie in the previously mentioned dependence of the emissivity on the temperature and density of the plasma. As we have constant pressure setups, higher density structures have lower temperatures, and the other way around. Thus, the brightest structures in the emission, which ultimately dominate in line ratio determination, have a specific density corresponding to the temperature of the peak in emissivity. The seismological results for the three different line ratios are listed in Table~\ref{densratio}.

\subsection{Isothermal vs. non-isothermal model}

In this study, we focus on equilibrium models with a straight and homogeneous magnetic field, which implies a multithermal plasma in the presence of density structuring, in order to maintain pressure balance. However, as the degree of (magnetic, thermal) structuring in the solar corona is largely unknown, a study of its potential effects on seismology diagnostics are needed. Therefore we employ an isothermal model, described in subsection 2.2, having the same density distribution as the $N=50$ model of Fig.~\ref{dynamics} to test the accuracy of the seismological method. Note that while in the multithermal model there is a homogeneous magnetic field of $5\ \mathrm{G}$, in the isothermal run we have an inhomogeneous magnetic field with an average magnitude of $\approx 5\ \mathrm{G}$.  First, we look at the effect of an isothermal setup on the density diagnostic: we obtain an average value of $6.81 \cdot 10^{-13}\ \mathrm{kg\ m^{-3}}$, close to double of the input average density. An explanation would be that, the structures being of the same temperature, the relative emissivity is depending only on the density profile, thus the brightest structures contributing to the diagnostic are the densest, offsetting the average to a higher value than in reality. The seismologically inferred magnetic field value for the $1.03\ \mathrm{MK}$ isothermal case is $6.06 \substack{+2.4 \\ -2.0}\ \mathrm{G}$, slightly higher than the multithermal estimates, and with larger spread, mainly caused by the uncertainty on the measured density. 

\subsection{Effect of instrument resolution}

Currently, the most suitable instrument to provide wave measurements for dynamic coronal seismology is CoMP \citep{2008SoPh..247..411T}, with a spatial resolution (pixel size) of 4.5 arcsec ($\approx 3.26\ \mathrm{Mm}$) and a cadence of 29 s. Therefore, we wish to study the dependence of seismological accuracy on the instrument resolution. For this, we artificially degrade the spatial and temporal resolution of the forward-modeled results, using the \texttt{rebin} and \texttt{congrid} function of \texttt{IDL}. The results for different resolutions are listed in Table~\ref{restable}.

\begin{table}[h!]
  \caption{Results of the seismological determination of the average magnetic field for different spatial resolutions and cadence, for the N=50 case. The meaning of errors and percentages is the same as for Table~\ref{table}.}
  	\centering
  	\begin{tabular}{|c|c|c|c|c|c|c}
  		\hline\hline
  		Spatial resolution (km) & Cadence (s)& $\langle |\mathbf{B}_{seism}| \rangle |\ (\mathrm{G})$ & $ |\mathbf{B}_{input}|\ (\mathrm{G})$ \\ 
  		\hline
  		default ($\approx 40$ km) & default (10 s) & $5.14\substack{+0.6 \\ -0.7}$ (2.8\%) & \multirow{7}{0em}{5} \\
  		312 & 10 &$5.14\substack{+0.6 \\ -0.7}$ (2.8\%)&\\
  		312 & 30 &$5.13\substack{+0.6 \\ -0.7}$ (2.6\%)&\\
  	   1250	& 10 &$5.06\substack{+0.5 \\ -0.7}$ (1.2\%)&\\
  	   1250 & 30 &$5.09\substack{+0.6 \\ -0.7}$ (1.8\%)&\\
  	   2500 & 30 &$5.09\substack{+0.4 \\ -0.6}$ (1.8\%)&\\
  	   5000 & 30 &$5.24\substack{+0.4 \\ -0.5}$ (4.8\%)&\\
  		\hline
  	\end{tabular}
  	\label{restable}
\end{table}

As we can see, the resolution degrading is not significantly affecting the inferred magnetic field value in this study. This is understandable as the interpolation to a coarser grid is similar to the averaging procedure used to obtain the inferred magnetic field value. The insensitivity of the inversion to the resolution and cadence is reassuring, as we expect sub-resolution fine structuring in the corona. Obviously, in real observations it is desirable to have as high a resolution as possible, in order to resolve the corona (and through seismology the magnetic field) to a high detail. \par

\subsection{Influence of the chosen spectroscopic line}
\label{spectroline}
The emissivity in a specific spectral line has a unique dependence on the density and temperature of the plasma. This implies that coronal structures appear in general differently in different spectral lines, which could influence the measured phase speeds from the spectroscopic Doppler images. As previously mentioned, in this study we use Fe XII and Fe IX lines, mostly for their popularity, to get synthetic spectroscopic images. Above we emphasized the importance of the interplay between emissivity and different parameters on the accuracy of the seismological calculation of the magnetic field. Therefore, different functions for line emissivities will lead to variable effects on the accuracy. Here we intend to study the influence of the chosen spectroscopic line on the measured phase speeds (via Doppler shifts). We keep the same lines for the line ratio study for density, as in the present range of densities the selected lines are the most viable choice. The multithermal $N=50$ setup was studied, and the apparent phase speeds were determined for the chosen four spectroscopic lines. Note that in fact the largest difference is between the Fe IX 171.073{\AA} and Fe XII lines, as they originate from different ionization states. The results of the comparison can be found in Table~\ref{spectable}.

\begin{table}[h!]
  \caption{Results of the seismological determination of the average magnetic field for different spectroscopic lines, for the N=50 case. The measured phase speed is also compared to the input average Alfv\'en speed. The meaning of errors and percentages is the same as for Table~\ref{table}.}
  	\centering
  	\begin{tabular}{|c|c|c|c|c|c|c}
  		\hline\hline
  		Spectral line ({\AA}) & $\langle {V_A}_{seism} \rangle (\mathrm{km\ s^{-1}})$ & $ \langle {V_A}_{input} \rangle (\mathrm{km\ s^{-1}}) $ & $\langle |\mathbf{B}_{seism}| \rangle\ (\mathrm{G})$ & $ |\mathbf{B}_{input}|\ (\mathrm{G})$ \\ 
  		\hline
  		default (Fe XII 195.119) & $649 \pm 30$ (11.2\%) & \multirow{4}{2em}{731} &$5.14\substack{+0.6 \\ -0.7}$ (2.8\%)& \multirow{4}{0em}{5} \\
  		Fe XII 186.887 & $635 \pm 30$ (13.1\%)& & $4.99\substack{+0.6 \\ -0.7}$ (0.2\%)&\\
  	    Fe XII 193.509 & $637 \pm 30$ (12.9\%)& & $5.00\substack{+0.6 \\ -0.7}$ (0.0\%)&\\
  	    Fe IX 171.073  & $627 \pm 40$ (14.2\%)& & $4.92\substack{+0.7 \\ -0.8}$ (1.6\%)&\\
  		\hline
  	\end{tabular}
  	\label{spectable}
\end{table}

\section{Discussion}

In section~\ref{results} we looked at the influence of various parameters on the accuracy of the seismological method for determining the magnetic field. These important parameters are either not known for the solar corona, such as the structuring, or depending on the instrument choice, such as spectral lines and resolution. Obviously, a study exhausting all the possible combinations of various parameters is not possible. However an ensemble view on how these parameters influence the accuracy of seismology was provided, in tabular form. \par
In some cases, we also list the measured average phase speeds, along with the input value of the average Alfv\'en speed. Note that the difference between these values does not reflect a measurement error: the real phase speed of the Alfv\'enic waves is not necessarily the average Alfv\'en speed, but depends on the various parameters studied, related to the structuring, in an essentially not computable manner. The difference rather reflects the performance of the average Alfv\'en speed as an inversion tool: the greatest difference found is $\approx 21 \%$, which, although without a proper reference point, is satisfactory having in mind previous errors on seismological estimates. As previously stated, the average Alfv\'en speed was used as the inversion tool in previous seismological studies of Alfv\'enic waves, mostly because of its simplicity. This naturally induces errors already at the theoretical stage of seismology. However it is arguably the best option available. This constitutes the theoretical limitation for the seismology of Alfv\'enic waves. \par
On the observational part, it is a well-known limitation that CoMP-like wave observations can only allow us to infer plane-of-sky (POS) magnetic field values through seismology \citep{2007Sci...317.1192T}: an angle $\alpha$ between the LOS and propagation direction leads to a ratio between the apparent POS phase speed and the real phase speed of $\sin(\alpha)$. In this sense, the inferred magnetic field magnitudes from real observations represent a minimum value. Note that we choose our LOS perpendicular to the propagation direction, therefore this does not apply in our case. Furthermore, the optically thin property of the coronal plasma means that emission is integrated along the LOS, leading to the loss of three-dimensional information and averaging.
\par
Judging by the accuracy of the inferred magnetic field values, the greatest source of error and spread for seismology seems to be the lack of a reliable average density diagnostic using the line ratio method.  This does not originate from an inherent error in the density diagnostic, as demonstrated by the measurements on homogeneous plasma (except the known variation with temperature), but from the previously described line-integrated emission in the presence of structuring. In the inhomogeneous setups, the average density is mostly overestimated, e.g. by a factor of $\approx 2$ for the isothermal $N=50$ setup, and also underestimated, e.g. by a factor of $\approx 2$ for the $N=250$ setup. The isothermal $N=50$ setup appears to perform the worst in inferring the average magnetic field value, with a $\approx 20\%$ error. \par
Furthermore, a general point has to be made on applying seismology to observations: unlike our ideal, simulated forward-modeled images, real observations present imperfect instrumentation and noise that could potentially reduce the seismological method's accuracy. In particular,  the line ratio method's usability is limited to plasma temperatures around the peak formation temperature of the respective ion. Emission from plasma at a substantially different temperature would be too weak and noisy for the ratio to be accurately measured, leading to an additional, potentially large source of error for density estimates. A possible solution to this problem is to use one of the lines of the chosen line ratio as the spectroscopic line used to measure the waves, or the use of a line ratio with formation temperature in the vicinity of that of the spectroscopic line, with the latter being the case in the present study: this would probably guarantee that if the waves are resolved (i.e. good signal-to-noise ratio), then line ratio measurements are possible as well, therefore this source of error would not play such a big role in density measurements. This is an important point which needs to be addressed in follow-up studies. \par
On a positive ending note, degrading the resolution or cadence does not seem to affect much the accuracy of the inferred values, nor does the choice of the spectroscopic line used, at least in the current setup and in the limited parameter space studied. This implies that sub-resolution fine structure in real observations might not significantly affect the accuracy of the inversions.  

\section{Conclusion} 
               
Coronal seismology is allowing for the measurement of elusive physical parameters of the coronal plasma, by comparing observed wave properties to MHD wave theory. It was successfully applied in the recent past to single events, and very recently to the ever-propagating transverse, Alfv\'enic waves. These ubiquitous waves could lead to the exciting possibility of dynamic coronal seismology, meaning continuous inversions of physical parameters, especially magnetic field magnitudes. These continuous inversions in time would allow for the study of e.g. the magnetic field evolution over an active region, with promising impact in many areas of solar physics and space weather. In this sense, assessing the capabilities of dynamic coronal seismology is an important step. As many important parameters directly determining the properties of Alfv\'enic waves in the solar corona are unknown, such as the structuring of the plasma across the magnetic field lines, relying on one type of model, e.g. that of cylindrical flux tubes, is unsatisfactory. We use three-dimensional models with various degrees of random density and temperature inhomogeneity, representing a thin, elongated column of coronal plasma. We then launch propagating waves mimicking the observed properties (periods, amplitudes) of ubiquitous Alfv\'enic waves. To be able to perform the same type of data analysis as performed on the observational data, we create synthetic spectroscopic images of the resulting dynamics in different spectral lines. The resulting Doppler time series are used to measure the phase speed of the propagating waves. The phase speed, supplemented by density diagnostics using the line ratio method allows for the seismological inversion of the magnetic field value, which is compared to the input value for the assessment of the method's accuracy. In previous studies, as well as here, it was assumed that the theoretical value of the phase speed of Alfv\'enic waves is the average Alfv\'en speed, which serves as the inversion tool for seismology. We provide a verification of this choice, and find that it appears satisfactory having in mind the lack of information on coronal structuring. We show that the spread in the inferred magnetic field value mostly comes from the problematic measurement of average densities. Despite this drawback, the inferred average magnetic field values are, in all different setups, within $\approx 20\%$ of the input value. The aim of this study is to assess the method rather than prove or disprove it, which is not possible without having a concrete error goal in mind.  Still, the seismological method appears to be robust and leading to encouraging magnetic field estimates, considering the wide range of different setups employed here. \par
This is a first study aimed at testing the coronal seismology of ubiquitous propagating Alfv\'enic waves. As a follow-up, we envisage more realistic tests, e.g. simulating ubiquitous waves in magnetic field geometries reminiscent of active regions, including full hydrodynamic evolution, and incorporating noise. Such studies would possibly further verify and strengthen the capabilities of dynamic coronal seismology.

\begin{acknowledgements} The authors would like to thank the referee for the valuable comments which helped to improve the manuscript. N.M. acknowledges the Fund for Scientific Research-Flanders (FWO-Vlaanderen). T.V.D. was supported by the GOA-2015-014 (KU Leuven) and the European Research Council (ERC) under the European Union's Horizon 2020 research and innovation programme (grant agreement No. 724326). Inspiration for this research was found during an ISSI Bern workshop organized by R. Morton and G. Verth entitled ``Towards Dynamic Solar Atmospheric Magneto-Seismology with New Generation Instrumentation" \end{acknowledgements}

\bibliographystyle{apj} % style aa.bst
\bibliography{../Biblio}{} % \begin{tiny}

\end{document}